\documentclass[ajp,showpacs]{revtex4}
\usepackage{graphicx}

\newcommand{\be}{\begin{equation}}
\newcommand{\ee}{\end{equation}}
\newcommand{\beqa}{\begin{eqnarray}}
\newcommand{\eeqa}{\end{eqnarray}}
\newcommand{\nonu}{\nonumber}

\begin{document}
  
 \par \section*{\bf{COMMENT:}  Numerical calculation of thermal noise-voltage in a Josephson junction  of finite capacitance [Phys. Lett. \bf{31A}, 314 (1970)]} 
 \medskip
  \begin{center} 
  Vinay Ambegaokar \\
  Department of Physics,  Cornell University, Ithaca NY 14853\\
 ( va14@cornell.edu)\\
 \end {center}

In the cited paper by the late J. Kurkij\"arvi and myself (KA) a molecular dynamics method is used to simulate the thermal agitation of the phase variable in a current carrying Josephson junction.  Since the method is general and may be useful in other contexts, it is briefly described with some details added.
 
 Consider a classical particle of mass $M$, position $X$, and momentum $P$ performing a one-dimensional thermal Brownian motion in a potential $U$.  The process may be described by the Langevin equations
 \be
 \dot X = P/M~; ~~\dot P = -{dU\over dx} - \eta P + L(t),
 \label{eq:Le}
 \ee
 where dots indicate derivatives with respect to the time $t$, and the symbols not yet defined are $\eta$ the average dissipation and $L(t)$ the fluctuating force.  It is known that to lead to thermal equilibrium at temperature $T$ (in energy units) the fluctuating force must have the autocorrelation function $\langle L(t) L(t')\rangle = 2\eta M T \delta( t - t')$, where the brackets indicate an appropriate average.
 
 The basic idea of the method introduced in KA is to simulate the fluctuating force $L(t)$ by random impulses describing elastic collisions with an ideal gas of light particles of mass $m$, assumed to be always in equilibrium.  From the collision dynamics one calculates that random impulses 2$p_i$ drawn from a set distributed  according to $g(p) = (|p|/2mT) \exp(-p^2/2mT),~ -\infty < p < \infty$, occurring  at random time intervals $t_i$ from a set distributed according to $f(t) = \nu \exp(-\nu t), ~0< t <\infty$, are required.  With a given choice of $m~\ll M$, $\nu$ must be chosen to satisfy $\nu = \eta M/4 m$, this being the relation between the mean frequency of impact $\nu$ and the damping constant $\eta$.
 
 Although the algorithm outlined in the last paragraph follows so directly from the physically motivated model that it cannot be wrong, it is worth verifying that the autocorrelation function of $L(t)$ so produced is indeed correct.  
 
 In the model $ L(t) = \sum_{i=1}^\infty 2p_i\delta (t - T_i), ~T_i=\sum_{j\le i}~  t_j$.  It follows that
 \be
\langle L(t) L(t')\rangle = \langle4 p_1^2\rangle \times \langle\delta(t - t_1) + \delta (t - t_1 -t_2) +\delta (t-t_1 -t_2 -t_3) + \delta(t-t_1-t_2-t_3 -t_4) +  \dots\rangle\times \delta(t-t').
\label{eq:Lem} 
\ee 
\par The fact that $\langle p_i p_j\rangle =0$, for $i\ne j$, has been used to keep only diagonal terms in a double sum.  In Eq.~(\ref{eq:Lem}), the first term (the average squared impulse)  is independent of its index and of time:  it has been taken out of the time-average.  Its value from the given distribution is $8 m T$.  

The average of the sum of $\delta$-functions is $\nu$.  This can be seen both intuitively and formally.  The intuitive argument is that, since the mean time interval between the $\delta$-functions is $\nu ^{-1}$, the integral of the average over a time $N/\nu$ is $N$. 

 The average can also be done exactly term by term.  The first term is simply
 \be
 \langle\delta(t-t_1)\rangle = \int_0^\infty d t_1~ \nu ~e^{-\nu t_1} \delta (t-t_1) = \nu~ e^{-\nu t}.
 \label{eq:first}
 \ee
 For the last term shown the average is
 \beqa
  A_4 \equiv \langle\delta(t-t_1-t_2-t_3 -t_4)\rangle& = &\int_0^\infty dt_2\int_0^\infty dt_3\int _0^\infty dt_4 ~\nu^4 e^{-\nu t} \Theta (t-t_2-t_3-t_4) \nonu\\ &=& \int_0^t dt_2 \int_0^{t - t_2} dt_3 \int_0^{t-t_2-t_3} dt_4 ~{\nu}^4 e^{-\nu t} = {1\over 3!}~ {\nu}^4~ t^3~e^{-\nu t}.
 \label{eq: f}
  \eeqa
  Above, the first equality comes from using the $ \delta$-function to collapse the $t_1$ integral, the step function $\Theta$ imposing the requirement that $t_1$ be positive. 
  
  [The average $A_4 $ may also be evaluated using Fourier transforms and a contour integration:
  \beqa
  A_4 &=& \int_{-\infty}^{+\infty} {d \omega\over 2\pi} e^{-i \omega t} [{\nu \over \omega +i\nu}]^4\nonu\\
  &=&-i\nu^4{1\over 3!}  {d^3\over d \omega^3} e^{-i\omega t}\vert_{\omega =-i\nu} = \nu~ {1\over 3!}(\nu t)^3 e^{-\nu t}.]
  \eeqa

 As illustrated, the successive terms in the sum of $\delta$-functions, when averaged over the random time intervals, produce an infinite series for $\exp (\nu t)$ which exactly compensates for and removes the decaying exponential in Eq.~(\ref{eq:first})!  The intuitive argument that the average of the sum of $\delta$-functions is equal to $\nu$ is thus formally verified.
  
 Inserting these evaluations into  Eq.~(\ref{eq:Lem}) one obtains
  \be
  \langle L(t) L(t')\rangle = 8 m T \nu\delta(t-t') = 2\eta M T\delta(t-t').
  \ee
  
  Note that in the final result the mass $m$ of the bath particles no longer appears, but that the temperature of the bath has been communicated to the Brownian particle.
  \section*{Acknowledgement}  This comment is dedicated to Ulrich Eckern on the occasion of his 60th birthday in gratitude for our several happy and successful collaborations. 
  
   \enddocument